\documentclass[aps,prb,amssymb,amsmath,showpacs,superscriptaddress]{revtex4}
\usepackage{graphicx}
\usepackage{longtable}

\begin{document}
\title{Experimental detection of quantum information sharing and their quantification in quantum spin systems}

\author{Diptaranjan Das}

\author{Tanmoy Chakraborty}

\author{Harkirat Singh}

\author{Chiranjib Mitra}
\email{chiranjib@iiserkol.ac.in}

\affiliation{Indian Institute of
Science Education and Research (IISER) Kolkata, Mohanpur Campus, PO:
BCKV Campus Main Office, Mohanpur - 741252, Nadia, West Bengal,
India}

\begin{abstract}

We study the macroscopic entanglement properties of a low dimensional quantum spin system by investigating its magnetic properties at low temperatures and high magnetic fields. The temperature and magnetic field dependence of entanglement from the susceptibility and magnetization data, comparing the experimental extraction with theoretical estimates are given. Extraction of entanglement has been made possible through the macroscopic witness operator, magnetic susceptibility. The protocol followed in doing so has been outlined in some recent work. Various plots of entanglement comparing the experimental extraction with theoretical estimates are given. Quantitative comparison between concurrence and entanglement witness is given for both the theoretical and experimental results. Theory and experiments match over a wide range of temperature and field. The spin system studied is a chain, which exhibits dimerisation and yields fascinating entanglement properties when the temperature and magnetic field are varied. These spin systems exhibit quantum phase transition (QPT) at low temperatures, when the magnetic field is swept through a critical value. We show explicitly for the first time, using tools used in quantum information processing (QIP), that quantum phase transition (QPT) can be captured experimentally using canonically conjugate observables. Macroscopically, quantum complementarity relation clearly delineates entangled states from separable states across the QPT. We have estimated the partial information sharing in this system from our magnetization and susceptibility data. The complementarity relation has been experimentally verified to hold in this system.

\end{abstract}
\maketitle
\section{Introduction}

Quantum computation and information has had tremendous appeal to researchers, both on the theoretical and experimental fronts \cite{NielsenChuang, Vlatko}. The most important element for quantum protocols is the quantum correlation arising from entanglement \cite{ZPhys, EPR_Paper, speakables}. Entanglement in quantum condensed matter systems has been extensively studied from the perspective of Quantum Information Processing for the last decade. Various measures of entanglement, and its quantification are now available \cite{Vlatko1}. Recently, there has been significant development in quantification of entanglement, especially in spin systems, where extraction of entanglement has been made possible through macroscopic witness operators like magnetic susceptibility \cite {Vlatko2, Rappoport, Souza, Tribedi}. Spin chains have been proposed for possible applications in making components like quantum wires which can be used to connect the various quantum registers or gates in a quantum computer. The study of entanglement characteristics of spin chains is important for their application in implementation of quantum communication protocols \cite{tele, dense} as well as for their potential application in constructing hardware components of a quantum computer. This has led to study of different properties of Heisenberg spin chains that would be used for computations as well as quantum communications \cite{bose02}. Spin systems, with their finite dimensional Hilbert spaces are ideal for studying the dynamics of entangled states. For this purpose the Heisenberg spin Hamiltonian with its exchange interaction is particularly instructive \cite{divizenzo, Bose1, sriram}. Since for the purpose of teleportation and other quantum protocols, maximally entangled channels provide quantum information transfer with very high fidelity, it is important that one should be able to quantify the entanglement of a particular system or channel \cite{durga1, durga2}.

In the present work, we have studied the magnetic properties of spin chains, which exhibits dimerisation and yields fascinating entanglement properties when the temperature and magnetic field dependence of the magnetic susceptibility is studied. We have extracted the temperature and magnetic field dependence of entanglement from the susceptibility and magnetization data. These spin systems exhibit quantum phase transition (QPT) at low temperatures, when the magnetic field is swept through a critical value. We have also obtained signatures of QPT through our witness operators. Here we have estimated the information sharing from our magnetization and susceptibility data, which are known to be complementary observables.

 Specifically, we study the entanglement in copper nitrate ($Cu(NO_3)_2 \times 2.5 H_2O$) using magnetic susceptibility as marcoscopic entanglement witness, where we report the variation of entanglement as a function of temperature as well as magnetic field. This system exhibits dimerisation and presents us the opportunity to study quantum phase transition (QPT) owing to a reduction in the relevant degrees of freedom, enabling one to capture the dynamics in a reduced Hilbert space. Due to this reduction, there is a clear crossover of the ground state and the first excited state as a function of magnetic field at low temperatures. We capture essential features of quantum phase transition through suitable quantum information processing tools. In this system the $Cu^{2+}$ has an unpaired, localized electron. It is known to be an alternating dimer spin chain, where the interaction between alternating spins (say spin number 2 and 3) is much less than that between the other two spins (spin number 1 and 2). This paper has been divided into two major sections, one devoted to study of entanglement in this system and the other is devoted to the study of quantum phase transition using the tools of quantum information processing.

 In the next section, we start with reviewing the entanglement property of spin chains, especially pertaining to its quantification through macroscopic entanglement witness in the form of susceptibility \cite {Vlatko2}. Subsequently we proceed to the magnetic properties of dimerised spin system, for explicitly demonstrating the effect of temperature and magnetic field on entanglement. In the following section, we present the high field magnetization data, especially pertaining to the observation of QPT. It is observed in the present case that a QPT separates the entangled and non-entangled Hilbert spaces.

\begin{figure}[htb]
\begin{center}
   \includegraphics[width=16.0cm]{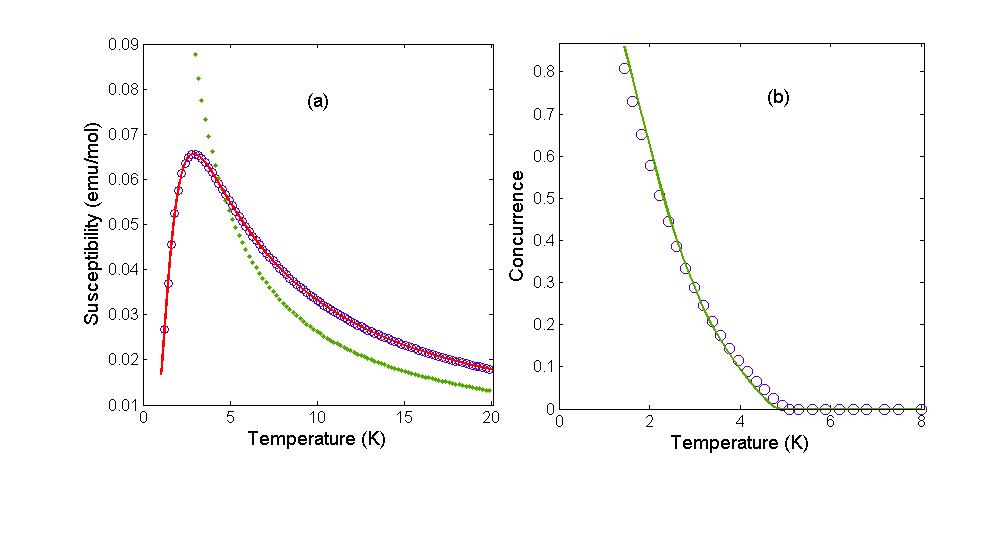}
\end{center}
 \caption{(a) Susceptibility data (circles) of $Cu(NO_3)_2 \times 2.5 H_2O$ and fit (solid line) to Eq. \ref{susceptibility}, corresponding to the alternating dimer model (Eq. \ref{susceptibility}). The dotted line marks out the separable regime
 (Eq. \ref{macrowitness}) from the entangled regime (see text). (b) Extraction of entanglement (concurrence) from the susceptibility data
 of copper nitrate using Eq. \ref{suscept-concurr}. The solid line is the theoretical fit to this experimentally extracted data using Eq. \ref{temperature}.}\label{copper-data}
 \end{figure}

\section{Entanglement in spin chains}

We briefly review the entanglement properties of different quantum states of a spin chain in this section. The Hamiltonian of a spin chain with nearest neighbour interaction, is given by,
\begin{equation}
\mathcal{H}=J \sum_{i} \vec{S_{i}} \cdot \vec{S}_{i+1} + B\sum_{i}
S_{i}^{z} \label{Hamiltonian}
\end{equation}
where, the spins are given by (in the unit of $\hbar$=1), $\
S_{i}=\frac{\sigma_{i}}{2}; i=1,2,3$ are the three Pauli spin
matrices mentioned earlier.

The magnetic field is assumed to be applied along the $\hat{z}$ direction. The net magnetization would be defined as the expectation value of $\sigma_{3}$, the eigenvectors of which are
$\vert\uparrow \rangle= \left(\begin{array}{c}
      1 \\
      0 \\
      \end{array}\right)$ and $\vert\downarrow \rangle=
\left(\begin{array}{c}
      0 \\
      1 \\
      \end{array}\right)$.

Since most of the entanglement measures currently available are for bipartite systems, it is worth exploring how one can extract the bipartite entanglement from a given susceptibility data for a spin chain. We consider a two spin system. The Hamiltonian for this two qubit system can be written as:
\begin{equation}
\mathcal{H}=J \vec{S_1} \cdot \vec{S_2} + B(S_1^z+S_2^z)=\\
\frac{J}{4} (\sigma_{1}^{x} \cdot \sigma_{2}^{x}+\sigma_{1}^{y}
\cdot \sigma_{2}^{y}+\sigma_{1}^{z} \cdot \sigma_{2}^{z} ) +
\frac{B}{2} (\sigma_{1}^{z} + \sigma_{2}^{z})
\label{Hamiltonian_two_qubit}
\end{equation}

The eigenvalues and the corresponding eigenvectors of this
Hamiltonian are: $E_1=\frac{J}{4}+B$; $\vert\uparrow \uparrow
\rangle= \left(\begin{matrix}

      1 \\

      0 \\

      0 \\

      0 \\

      \end{matrix}\right)$, $E_2=\frac{J}{4}-B$; $\vert\downarrow
\downarrow \rangle= \left(\begin{matrix}

      0 \\

      0 \\

      0 \\

      1 \\

      \end{matrix}\right)$, $E_3=\frac{J}{4}$; $\vert\psi^{+} \rangle =\frac{1}{\sqrt{2}}(\vert\uparrow \downarrow \rangle + \vert\downarrow \uparrow \rangle)= \frac{1}{\sqrt{2}}\left(\begin{matrix}

      0 \\

      1 \\

      1 \\

      0 \\

      \end{matrix}\right)$, $E_4=-\frac{3J}{4}$;
      $\vert\psi^{-}\rangle =
\frac{1}{\sqrt{2}}(\vert\uparrow \downarrow \rangle -
\vert\downarrow \uparrow \rangle)=
\frac{1}{\sqrt{2}}\left(\begin{matrix}

      0 \\

      1 \\

      -1 \\

      0 \\

      \end{matrix}\right)$,
      where $\vert\psi^{+}\rangle$ and $\vert\psi^{-}\rangle$ are the Bell states.

A ferromagnetic system ($J<0$), in the absence of external magnetic field doesn't display any entanglement in the ground state. This is
because the order parameter (magnetization) commutes with the Hamiltonian and there is no spin fluctuation at low temperatures. However, for an antiferromagnetic ground state, i.e., for systems exhibiting $J>0$, the order parameter, which is the staggered magnetization, does not commute with the Hamiltonian and thus there will be spin fluctuation, even at zero temperature. Hence, this system is entangled in absence of any external field. In fact, the ground state of an antiferromagnetic interaction ($J>0$) is the maximally entangled state ($\vert\psi^{-}\rangle$) for the above system, whereas, the ground state for $J<0$, is an equal mixture of the three fold degenerate triplet states and is a separable state.

In absence of external magnetic field the order parameter (staggered magnetization) doesn't commute with the Hamiltonian leading to non-trivial dynamics at zero temperature. Thus the order parameter can be used as an entanglement witness, an observable whose expectation value is non-zero for an entangled state and is zero for a separable state. Magnetic susceptibility has been shown to be a macroscopic entanglement witness \cite{Vlatko2}, where the authors show that susceptibility is a measure of entanglement, which is a non-local property of the spins. Its canonically conjugate observable, magnetization is a measure of the local spin properties and is therefore an observable complementary to the magnetic susceptibility. The magnetic susceptibility can be measured experimentally and it is possible to extract a measure of entanglement from the experimental data \cite{Vedral, IISc}. One of the first systems which exhibited entanglement through bulk measurements like magnetic susceptibility or heat capacity is the insulating magnetic salt LiHo$_x$Y$_{1-x}$F$_4$~\cite{gabriel}.

\begin{figure}[htb]
\begin{center}
   \includegraphics[width=10.0cm]{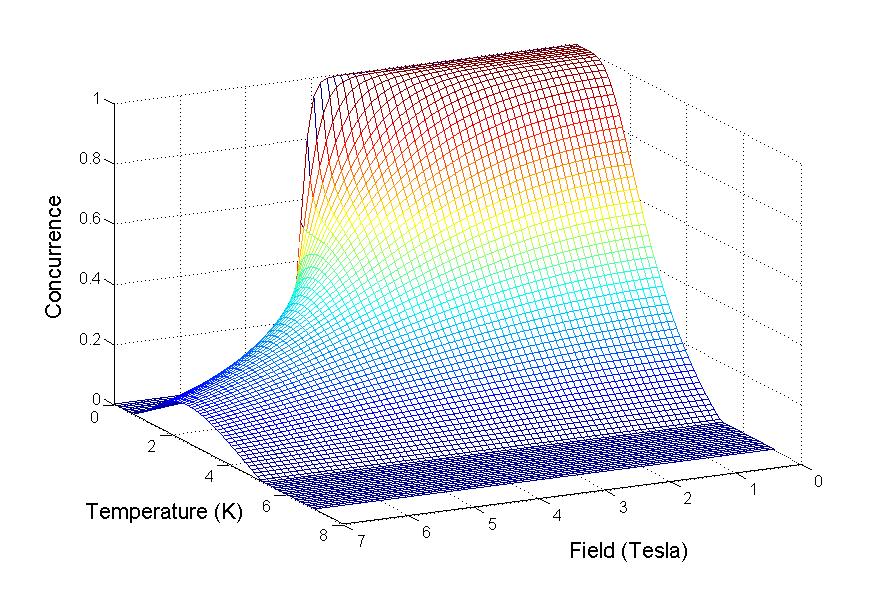}
\end{center}
 \caption{Theoretical value of concurrence as a function of magnetic field and temperature for a dimer model. The temperature and magnetic field values are in arbitrary units.}\label{conc-theor}
 \end{figure}

The magnetic susceptibility, $\chi$, can be defined as the field derivative of magnetization $M$ (in the applied field direction, which here is $\hat{z}$) in the limit $B\rightarrow0$, and the magnetization, M, is defined as the sum of the expectation values of the $\hat{z}$-component of individual spins, ${M}= \sum_{i=1}^{N}\langle \sigma_{z}^{i} \rangle $. The expression for susceptibility derived from an alternative dimer model is given as \cite{Vedral, IISc},

\begin{equation}
\chi= (\frac {\delta{M}}{\delta{B}})=\! (g^2 \mu^2_B / k_{B}T) \langle
M^2 \rangle_{B=0}\!=\!(g^2 \mu^2_B / k_{B}T) \sum_{i,j} \langle {S_i}^z
{S_{i+1}^z} \rangle \approx \\ (g^2 \mu^2_B N/ k_{B}T) [(1/4) + (\langle
\vec{S}_{1} \cdot \vec{S}_{2} \rangle)/3 ], \label{susceptibility}
\end{equation}

where, $N$ is the total number of spins per mole, $\mu_B$ is the Bohr magneton and $k_{B}$ is the Boltzmann constant.
It was derived using the isotropy of space, nearest neighbour interaction and the fact that $(\vec{S}_{1}+\vec{S}_{2})$=0 for an antiferromagnet.
One can evaluate $\langle \vec{S}^{1} \cdot \vec{S}^{2} \rangle$, from the bipartite Hamiltonian (Eq.
\ref{Hamiltonian_two_qubit}), for the $B=0$ limit, yielding,

\begin{equation}
\langle
\vec{S}^{1} \cdot \vec{S}^{2} \rangle= (\frac{-3}{4})
\frac{1-e^{-\frac{J}{kT}}}{1+3e^{-\frac{J}{kT}}} \label{sus_theor}.
\end{equation}

To extract the entanglement, one needs to consider the definition given by Wootters \cite{Wooters1}where it has been shown that concurrence is a good measure of entanglement. Concurrence is given by,

\begin{equation}
\mathcal{C}=max(0, {\sqrt{\lambda_1}-\sqrt{\lambda_2}-\sqrt{\lambda_3}-\sqrt{\lambda_4}}),
\label{concurrence}
\end{equation}
where $\lambda_1 \ge \lambda_2 \ge \lambda_3 \ge \lambda_4$, are the
eigenvalues of the operator,
\begin{equation}
\tilde{\rho}_{12}=\sigma_2\otimes\sigma_2 \rho_{12}^*
\sigma_2\otimes\sigma_2, \label{rho-tilde}
\end{equation}
such that, $\rho_{12}$ is the two particle reduced density matrix
and the asterisk denotes complex conjugation. It has been shown by
O'Connor and Wootters \cite{Connor}, that for an antiferromagnet,
the concurrence is given by,
\begin{equation}
\mathcal{C}=\frac{1}{2} \textrm{ max }[0,\frac{\vert U \vert}{N J }
-1] , \label{antiferromag}
\end{equation}
in absence of magnetic field. Here, $U$ is the free energy, which, by definition, is the expectation value of the Hamiltonian, i.e., $U= \langle \mathcal{H} \rangle $.
Thus, considering the system to be isotropic, the formula for concurrence, given in Eq.
\ref{antiferromag}, can be written as,
\begin{equation}
\mathcal{C}=2 \textrm{ max }[0,\langle \vec{S}^{1} \cdot \vec{S}^{2}
\rangle-(1/4)]. \label{correlation}
\end{equation}

Using Eq. \ref{sus_theor} we have,
\begin{equation}
\mathcal{C}=2 \textrm{ max }[0,
\frac{1-e^{-\frac{J}{kT}}}{1+3e^{-\frac{J}{kT}}}]
\label{temperature}
\end{equation}

From the condition of separability of two particle states and Eq. (\ref{susceptibility}) one
can obtain the concurrence in terms of susceptibility,

\begin{equation}
\mathcal{C}=\textrm{ max }[0, 1- \frac{(6k_{B}T\chi)}
{g^{2}\mu_{B}^{2}N}]. \label{suscept-concurr}
\end{equation}

Thus for a separable state, the magnetic susceptibility will satisfy,
\begin{equation}
\chi \geq \frac{g^2 \mu^2_B N}{kT} \frac{1}{6}. \label{macrowitness}
\end{equation}

\begin{figure}[htb]
\begin{center}
   \includegraphics[width=13.0cm]{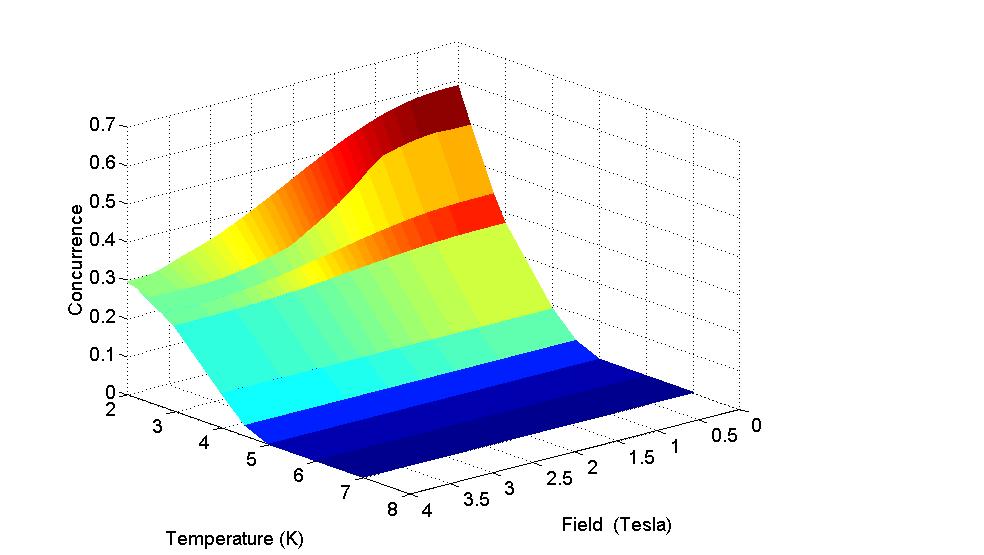}
\end{center}
 \caption{Experimental value of concurrence as a function of magnetic field and temperature for $Cu(NO_3)_2 \times 2.5 H_2O$. The magnetic field values are in Tesla and the temperature is in Kelvin.}\label{conc-data}
 \end{figure}

Entanglement witness, which in spin systems is susceptibility is a more general quantification of entanglement and hence could be applicable to a general spin chain, with spin - $\frac{1}{2}$, spin-1 or spin-$\frac{3}{2}$. However, concurrence is defined only for spin - $\frac{1}{2}$ system \cite{Wooters1} and this measure of entanglement has a striking match with magnetic susceptibility as a measure of entanglement. Here, the expression for concurrence given in Eq. \ref{suscept-concurr} is the entanglement witness which will be extracted from experimental data below. Higher spin systems will contain higher dimensional entanglement and thus the deviation from entanglement witness is more \cite{Souza1}. The system that we have considered for this paper is copper nitrate ($Cu(NO_3)_2 \times 2.5 H_2O$), an inorganic compound and is a spin - $\frac{1}{2}$ system. The sample was procured from Sigma Aldrich. $Cu(NO_3)_2 \times 2.5 H_2O$ has been shown to be an alternative dimer spin chain \cite{Xu}. The corresponding spin Hamiltonian would be of the type,

\begin{equation}
H=\sum_i (J_1 \mathbf{S}_{2i} \cdot \mathbf{S}_{2i+1} + J_2
\mathbf{S}_{2i+1} \cdot \mathbf{S}_{2i+2}), \label{alternate_dimer}
\end{equation}

representing pairs of spins that are alternately coupled by strong intra-dimer coupling $J_1$ and weak inter-dimer coupling $J_2 \!\approx \! 0.25~J_1$ \cite{Xu}. This system behaves like a chain of dimers or exchange coupled pairs which are themselves weakly coupled with each other.

Entanglement in $Cu(NO_3)_2 \times 2.5 H_2O$ has been reported earlier by Brukner $\it et~al$ \cite{Vedral}, who used the susceptibility data of Berger $\it et~al$ \cite{CuNO3} to extract entanglement. We have measured the temperature and magnetic field dependence of susceptibility of $Cu(NO_3)_2 \times 2.5 H_2O$ from 2~K to 20~K and from 0 to 4 Tesla. The temperature dependence of susceptibility in zero field is shown in fig \ref{copper-data}(a). All magnetic data was obtained using Quantum Design Magnetic Property Measurement System (MPMS).

%\begin{figure}[htb]
%\begin{center}
%   \includegraphics[width=10.0cm]{conc3D_theor.eps}
%\end{center}
% \caption{Theoretical value of concurrence as a function of magnetic field and temperature for a dimer model. The temperature and magnetic field values are in arbitrary %units.}\label{conc-theor}
% \end{figure}

\begin{figure}[htb]
\begin{center}
   \includegraphics[width=10.0cm]{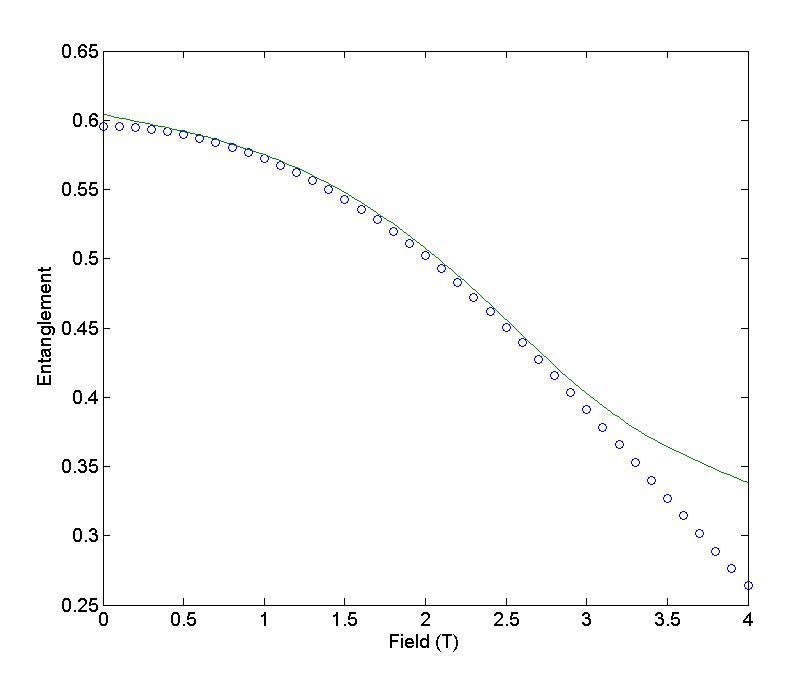}
\end{center}
 \caption{Experimental value of concurrence as a function of magnetic field at T~=~2K (open circles). The theoretical concurrence at T~=~2K is also plotted as a function of field (dashed lines).}\label{E_vs_B}
 \end{figure}

The zero field susceptibility data was fitted to the formula for susceptibility given in Eq. \ref{susceptibility} (see Fig.\ref{copper-data}(a)). The solid line shows the fitted curve. The data (open circles) and the fit (solid line) are in good agreement with each other. The dotted line shows the entanglement boundary given by Eq. \ref{macrowitness}; the entangled region is represented by the region towards the left of the dotted line. The copper spins exhibit entanglement up to the temperature where antiferromagnetic correlations persists, i.e. up to around 5 K. This is slightly more than the ordering temperature, which corresponds to the peak in the susceptibility curve. Fig. \ref{copper-data}(b), shows the extracted value of entanglement from the experimental data (open circles), which were extracted using the expression for entanglement witness (concurrence) given in Eq. \ref{suscept-concurr}. The theoretical fit (solid line) to this experimentally extracted data is also shown here. It was generated using the expression for concurrence given in Eq. \ref{temperature} for J=4 K. It can be seen in this figure that the concurrence vanishes above the temperature where the antiferromagnetic correlations vanish. Here too the theoretical fit (solid line) matches reasonably well with the experimentally extracted values of concurrence generated using the entanglement witness (fig. \ref{copper-data}(b)). Thus entanglement is, not only an entity which is quantifiable, but it also is something that can be measured experimentally.

The effect of temperature and magnetic field on the entanglement of a Heisenberg anti-ferromagnet, considering the dimer model has been dealt theoretically by Arnesen, Bose and Vedral \cite{Bose1}, wherein they consider the effect of temperature induced magnons on entanglement. In fig. \ref{conc-theor} we re-generated the theoretical 3D plot given in fig. 1 of Arnesen $et.~al.$ \cite{Bose1}, however, unlike theirs, which was plotted in arbitrary units, here the temperature and magnetic field has been scaled exactly to match with experimental data. The magnetic field values are in Tesla and the temperature is in Kelvin. The entanglement (concurrence) in this plot was generated for the dimer model, considering J~=~4~K. The experimental 3D plot of concurrence as a function of field and temperature is shown in fig. \ref{conc-data}. The visual difference between the theoretical and experimental 3D plots is due to the fact that the lowest temperature in the theoretical plot is zero Kelvin, and the maximum magnetic field is 7 Tesla, whereas the minimum temperature displayed in the experimental 3D plot is 1.8~K and the maximum field is 4 Tesla. Though we have collected magnetization data up to 7 Tesla, we have shown the experimental plot only up to 4 Tesla. This is partly because most of the relevant physics pertaining to this section is captured up to 4 Tesla, and partly because of some subtleties involved in the analysis of susceptibility data above 4 Tesla, which is mentioned in the next section. One can see the general pattern of decreasing entanglement with increasing temperature is consistent in both the theoretical and experimental plots, so is the decrease in entanglement with increase in magnetic field at the lowest measured temperature. This is due to the greater contribution from the separable singlet state in the statistical mixture as the temperature is increased at a particular field or as the magnetic field is increased at a particular temperature (see fig. \ref{conc-data}). The plot also shows that the entanglement vanishes at the same temperature (around T~=~5K) irrespective of the strength of magnetic field.

In fig. \ref{E_vs_B}, we have shown the plot of experimental value of entanglement as a function of magnetic field at T=2~K. Also shown in this figure is the theoretical entanglement as a function of field. As expected, this shows a gradual decrease from the maximum value. To understand this let us consider the following. At a finite temperature, say T~=~2K, the state is in a mixture of the four eigenstates, the thermal density matrix of this state is given as,

\begin{equation}
\rho=\frac{1}{Z} \{ \vert \phi^{-} \rangle \langle \phi^{-}\vert
e^{\frac{3J}{4}\beta} + \vert \phi^{+} \rangle \langle \phi^{+}\vert
e^{-\frac{J}{4}\beta} +  \vert \uparrow \uparrow \rangle \langle
\uparrow \uparrow \vert e^{-(\frac{J}{4}-B)\beta} + \vert \downarrow
\downarrow \rangle \langle \downarrow \downarrow \vert
e^{-(\frac{J}{4}+B)\beta}\},
\end{equation}

where, $\beta=\frac{1}{k_{B}T}$ and $Z=Tr(\rho)= e^{\frac{3J}{4}\beta}+e^{\frac{-J}{4}\beta}+e^{(\frac{-J}{4}+B)\beta}+e^{(\frac{-J}{4}-B)\beta}$.

\begin{figure}[htb]
\begin{center}
   \includegraphics[width=14.0cm]{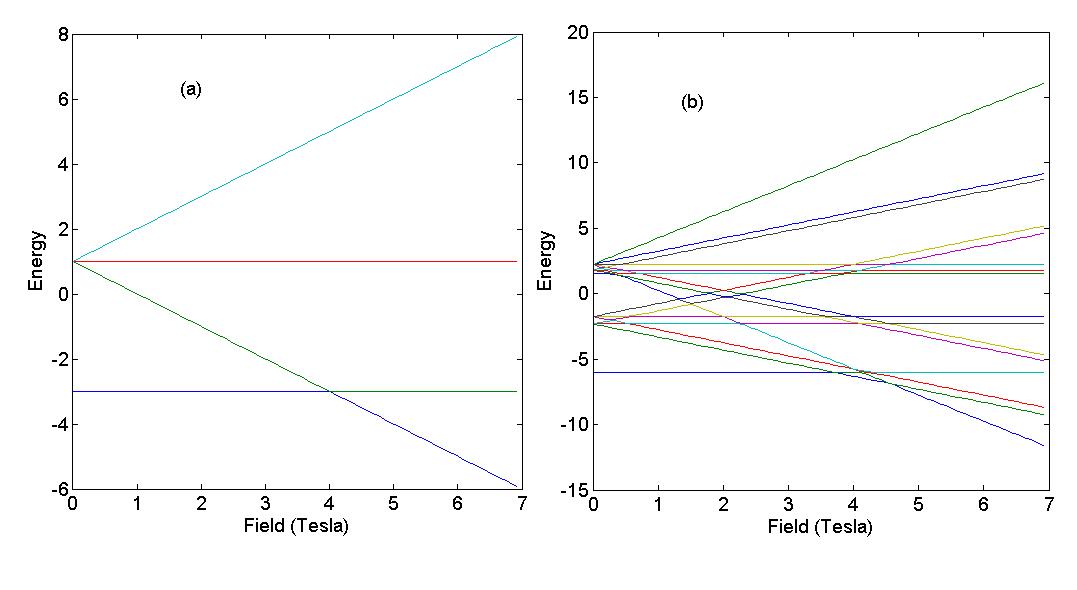}
\end{center}
\caption{(a) Energy eigenvalues vs magnetic field for a dimer system. (b) Energy eigenvalues vs magnetic field for an alternating dimer system.}\label{qptab}
\end{figure}

At finite (non-zero) temperatures the system is in a statistical mixture of all the eigenstates. At temperatures as low as 2K two lowest energy states, $\vert\psi^{-}\rangle$ and $\vert\downarrow \downarrow \rangle$, will have the dominant contribution. Thus, as the magnetic field is increased, the contribution from $\vert\downarrow \downarrow \rangle$, a separable state, increases gradually at the expense of contribution from $\vert\psi^{-}\rangle$, which is a maximally entangled state. As one keeps on increasing the magnetic field, the $\vert\downarrow \downarrow \rangle$ state ought to become cross the $\vert\psi^{-}\rangle$ state and become the lowest energy state above a critical field. This corresponds to the quantum phase transition (QPT) as can be seen from the theoretical 3D plot of concurrence (\ref{conc-theor}) as a function of field and temperature. Till the quantum critical point (QCP), the eigenstates do not change, though the eigenvalues would. At zero temperature the ground state would remain intact below QCP, but for finite temperature we would have a statistical mixture, skewed in favour of the two states mentioned above, closing the excitation gap between the two states. However, despite the closing of this gap, one can still see signatures of QPT at finite temperatures, and the detailed analysis is given in the next section.

In fig. \ref{E_vs_B}, we can see that the experimental variation of concurrence at a fixed temperature is in reasonably good agreement with the theoretical curves especially at low magnetic fields. The slight difference between the theoretical and experimental curves at high fields (above 3~Tesla) is probably due to slight difference in $J$ in the two scenarios. This discrepancy arises because the theoretical model considered by Arnesen $et~al$ \cite{Bose1} is a dimer model (with $J~=~4~K$) whereas $Cu(NO_3)_2 \times 2.5 H_2O$ is an alternative dimer system corresponding to Eq. \ref{alternate_dimer}, with $J_1~=~0.25~J_2$. In addition to that, the isotropy of the system i.e the fact that $\chi_x=\chi_y=\chi_z$ breaks down in the case of high magnetic field. However, since this is a dimerised system, leading to formation of singlet states, where the net $\overrightarrow{S}=0$, there is perfect SU(2) symmetry, as the system is described by isotropic Heisenberg model, one can use the formula for entanglement witness till a field of 4 Tesla for a system where $J\approx4K$.

\section{Quantum Phase Transition}

Quantum phase transition (QPT) in a quantum spin system takes place when a level crossing occurs between the ground state and excited states as one varies the magnetic field, such that an excited state becomes the new ground state \cite {Subir1}. This can happen either in a transverse field Ising model \cite {Subir1} or in isotropic Heisenberg model where dimerization occurs \cite {Subir2}. The point where the cross over takes place is called the quantum critical point. Behaviour of entanglement at the quantum critical point has been extensively studied
theoretically by several groups \cite{osborne, osterloh, vidal}. They have found that the
entanglement scales close to the quantum critical point (QCP). This means that the entanglement, which in turn
is related to the correlation function, is scale invariant close to the QCP. The QPT in dimer like
systems is characterised by crossing of energy eigenstates as a function of some parameter like the magnetic field.
This scenario is depicted in fig. \ref{qptab}(a) for a dimer system (two qubit system), where the state $\vert\uparrow \uparrow
\rangle$ crosses the ground state at some critical field and replaces it as the ground state, thus changing
the symmetry of the ground state. Thus the ground state crosses over from a maximally entangled state to a separable
state for a two qubit system. The exchange interaction for this system, $J/k_B~=~4~K$. This is chosen to suit copper nitrate.

\begin{figure}[t]
\includegraphics[width=0.6\columnwidth,clip]{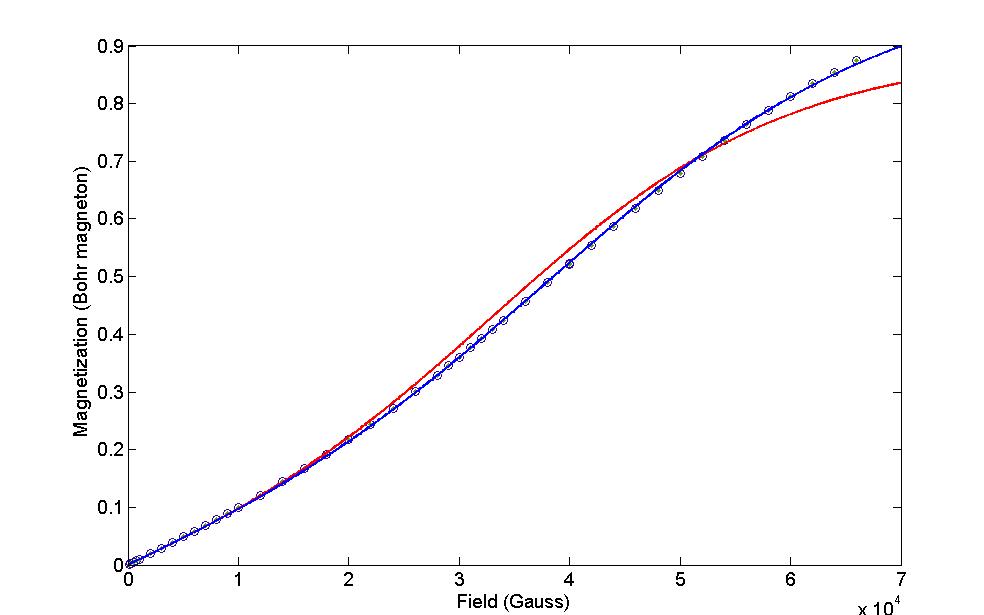}
\caption{Experimental data of magnetization collected at T~=~2~K for $
\mbox{$Cu(NO_3)_2 \times 2.5 H_2 O$}$\ (open circles) as a function of magnetic field and the fit to the theoretical curve (solid blue line)
derived using the alternating dimer model. The dotted line (red) is fit to the experimental data using a dimer model (Eq. \ref{magnetization}).
} \label{magnet_2k}
\end{figure}

\begin{figure}[t]
\includegraphics[width=0.6\columnwidth,clip]{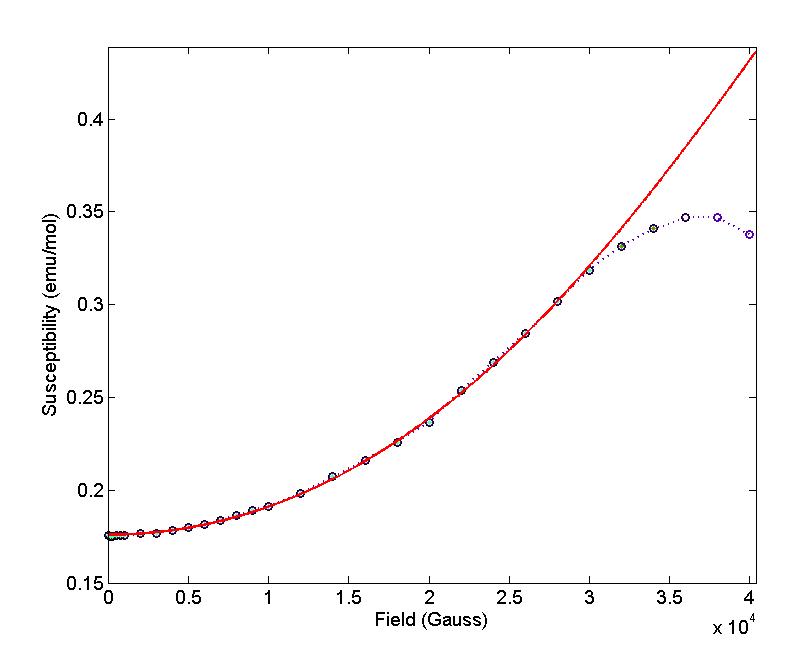}
\caption{Experimental data of magnetic susceptibility collected at T~=~2~K for $
\mbox{$Cu(NO_3)_2 \times 2.5 H_2 O$}$\ (open circles) as a function of magnetic field and the fit to the theoretical curve (solid line)
represented by Eq. \ref{susc_fld}. The dotted line is a guide to the experimental data.
} \label{susc_fit}
\end{figure}

The minimum amount of spins needed to fully describe an alternating dimer system is four, the dimension of the Hilbert
space being sixteen. The Hamiltonian for such a 4-qubit system is given by,

\begin{equation}
\mathcal{H}=J_1 \vec{S_1} \cdot \vec{S_2} + J_2 \vec{S_2} \cdot \vec{S_3} + J_1 \vec{S_3} \cdot \vec{S_4} + B(S_1^z+S_2^z+S_3^z+S_4^z),
\label{Hamiltonian_four_qubit}
\end{equation}

where $J_1$ and $J_2$ are the two exchange interactions, with  $J_1$ being significantly larger than $J_2$. The energy eigenvalues for
this Hamiltonian has been plotted as a function of field for $J_1/k_B~=~4~K$ and $J_2~=~0.25 \times J_1$. One can see that there are two crossovers
in the ground state, as is depicted in fig. \ref{qptab}(b). Thus the system will have a lower critical field and an upper critical field.

In a significant effort to understand the quantum correlations close to a QPT, concepts from quantum information processing, especially that of entanglement was invoked \cite{osterloh}. This was done especially to extract the extra correlations present in QPT, which do not have a classical counterpart. It was found that close to the critical point, the entanglement depends strongly on the magnetic field. Now since entanglement measures the non-local effect in the system, we also tried to capture the non-locality in the spin system close to QCP.

Here we present the detailed analysis of the magnetic field dependence of
magnetization and susceptibility. It has been reported by $Wiesniak~et.~al.$ \cite{Vlatko2}
that magnetization and susceptibility are complementary variables. Thus, whereas magnetization captures the local properties of spins,
susceptibility exhibits non-local features in a quantum spin system (as is evident
from its relation with correlation function). The non-local nature of entanglement
intuitively points towards the connection between entanglement and susceptibility
(c.f. Eq. \ref{suscept-concurr}) which in turn is connected with spin-spin correlation functions.

$Wiesniak~et.~al.$ \cite{Vlatko2} have derived a
macroscopic quantum complementarity relation, which using their notation is given as,

\begin{equation}
\label{comp} \underbrace{1-\frac{kT\bar{\chi}}{Ns}}_{non-local \mbox{ } properties} +
\underbrace{\frac{\langle \vec{M}\rangle^2}{N^2s^2}}_{local \mbox{ } properties} \leq 1.
\end{equation}

Again following their notation, let $Q \!\equiv\! 1-\frac{kT\bar{\chi}}{Ns}$ and
$P\equiv\frac{\langle\vec{M}\rangle^2}{N^2s^2}$, where $P$ describes
the local properties of individual spins, $Q$ the non-local
quantum correlations between spins. Since $Q$, through its dependence on the susceptibility
is proportional to two-point correlation function, its
positive value implies the existence of entanglement (c.f.
Eq.~(\ref{suscept-concurr})). When $Q=1$, the system is maximally entangled and in such a case, the local property of spins,
$P=0$. In the other extreme, when $P=1$ or the magnetization is a maximum, the local properties of the spins is
well defined at the expense of the non-local property, $i.e.$, $Q=0$. Thus, mathematically, the above relation
(\ref{comp}) describes partial quantum information sharing between local and non-local properties of spins. Following the work of $Wiesniak~et.~al.$, a similar complementary relation was also demonstrated theoretically by $Tribedi~et.~al.$ \cite{Tribedi1} for a spin 1 system.

At the heart of this inequality lies the fact that the spin operators, $\hat{S_x}$, $\hat{S_y}$ and $\hat{S_z}$, or the Pauli spin matrices (in case of a qubit) do not commute. The inequality, $\langle\sigma_x\rangle^2\!+\!\langle\sigma_y\rangle^2\!+\!\langle\sigma_z\rangle^2\!\leq\!
1$, where the average is taken over an arbitrary state, suggests that for a mixed state the spin lies inside the Bloch sphere, whereas for a pure state it will lie on the Bloch sphere. In the latter case, the equality relation holds as in such a case it is a pure state. For a particle with spin s, this inequality reads, $\langle S_x\rangle^2$ + $\langle S_y\rangle^2$ + $\langle S_z \rangle^2\!\leq\! s^2$. If we consider $\langle S_x^2 \rangle^2$+ $\langle S_y^2 \rangle^2$+ $\langle S_z^2 \rangle^2$, the value will always be more than $s^2$ and will in fact be $s(s+1)$. This extra value comes from the fact that $\hat{S_x}$, $\hat{S_y}$ and $\hat{S_z}$ do not commute. Thus, there is an uncertainly in determining the values of all three components of spin simultaneously. For a
composite system, this results in extra correlations which essentially is  non-local. For such a composite system, entanglement plays the role analogous to that of a one qubit coherently superposed pure state that lies on the Bloch sphere. Since, the susceptibility, $\chi$ $\propto$ $(\langle M^2\rangle$ - $\langle M\rangle^2)$, the second term being less than or equal to $N^2s^2$, N being the total number of spins, whereas the first term being proportional to $N^2~s(s+1)$, the inequality given in Eq. 6 of Wiesniak $et. al.$ \cite{Vlatko2} holds.

We have analysed our magnetization and susceptibility data extensively and tried to ascertain the experimental validity of this relation.
In what follows is a careful exposition of metrology and the validity of susceptibility measurements at high magnetic fields.

\begin{figure}[t]
\includegraphics[width=0.9\columnwidth,clip]{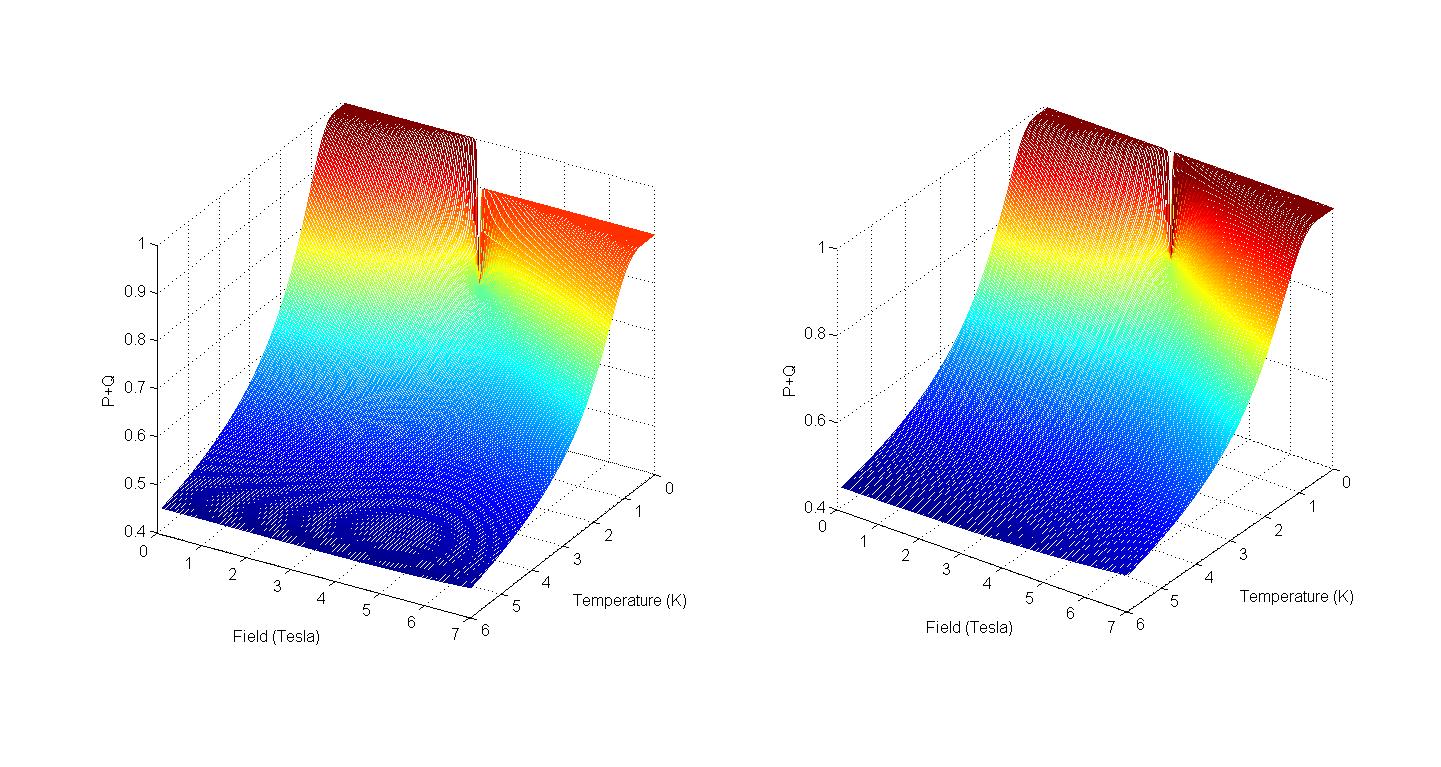}
\caption{(a) Plot of theoretical P+Q as a function of magnetic field and temperature using $\left(\sum_{i,j=1}^N \langle s_z^i s^j_z\rangle-\left\langle \sum_{i=1}^Ns_z^i \right\rangle^2 \right)$ for constructing the susceptibility (b) The same where susceptibility was constructed using $\left(\sum_{i,j=1}^N \langle s_z^i s^j_z\rangle\right)$.} \label{PQ_3D}
\end{figure}

The magnetization data for copper nitrate is given in fig. \ref{magnet_2k}.  Though the entire
field and temperature range has been shown in fig. \ref{conc-data}, here we focus on the data
set below the antiferromagnetic ordering temperature, where all the interesting features are occurring. The data was fitted to both the
dimer as well as the alternating dimer model.  Eq. \ref{magnetization} gives us the expression for a
dimer model.
\begin{equation}
M=Tr(\rho\cdot
S_z^{Total})=\frac{2sinh(2g\mu_BHk_BT)}{1+2cosh(2g\mu_BHk_BT)+exp(J/k_BT)},
\label{magnetization}
\end{equation}
where ``g" is the Land\'{e} g factor.

The fit to this expression is given in blue. We can see that the dimer
model is not a very good representation of the real spin system. The expression for
alternating dimer model was generated numerically and subsequently used to fit the
magnetization data. The expression is not explicitly shown here owing to its
cumbrous nature. We can see that it is a very good fit. The value of ``J'' obtained
from this fit to all the four data set shown in fig. \ref{magnet_2k} is around 4K. This only goes
to show that the $ Cu(NO_3)_2 \times 2.5 H_2 O$. is an alternating dimer system as
is also evident from neutron data \cite {Xu}.

\begin{figure}[t]
\includegraphics[width=0.7\columnwidth,clip]{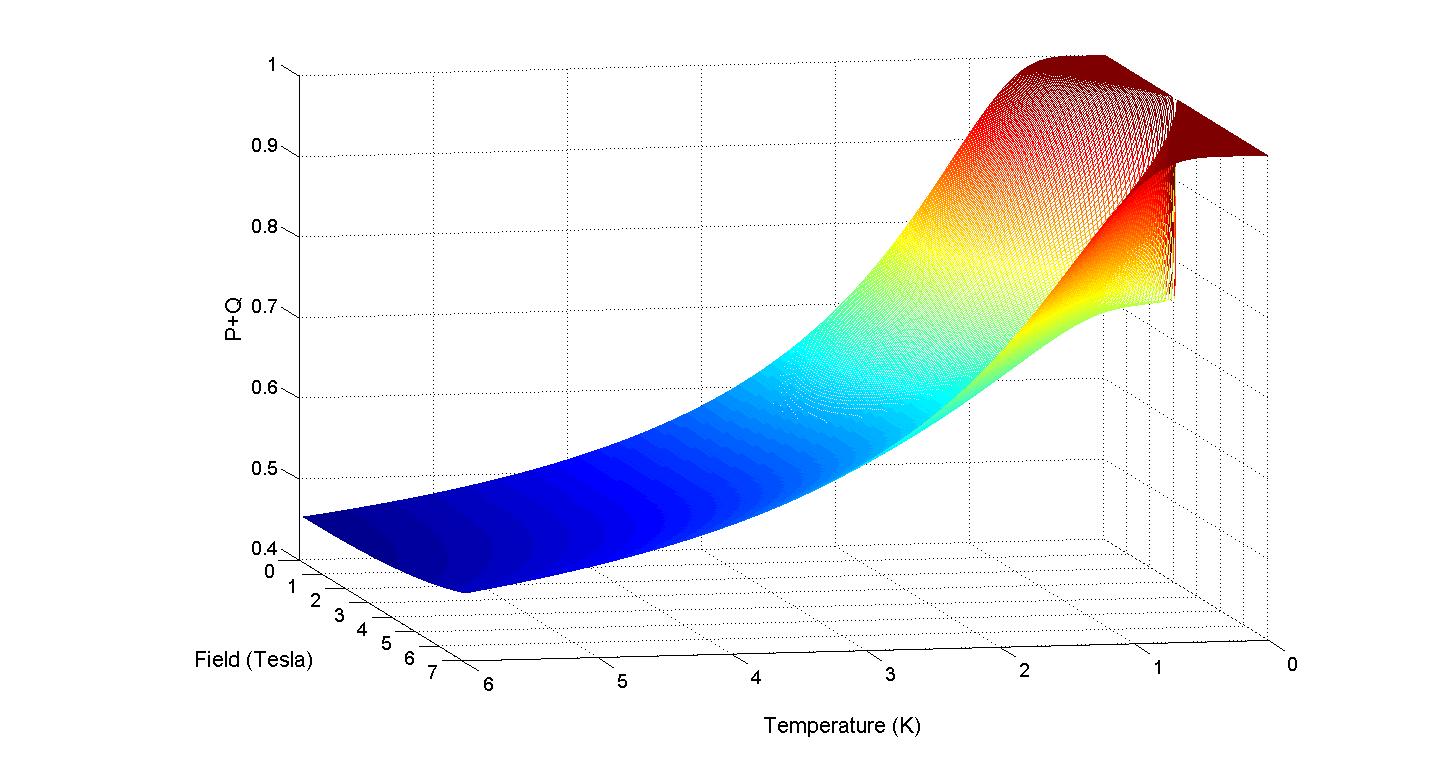}
\caption{Side view of the theoretical P+Q as a function of magnetic field and temperature shown in fig. \ref{PQ_3D}(b).} \label{P+Q_3}
\end{figure}

Hereafter, we analyze the susceptibility data as a function of magnetic field acquired at the minimum temperature (2 K), shown in fig. \ref{susc_fld}. The susceptibility data shows a down turn after 4 Tesla . Here it is imperative to mention a few lines about the nature of
measurement of generalized susceptibility employed in the MPMS or other magnetic
susceptibility measurement system.

The zero field magnetic susceptibility is defined as, $\chi_z\!=\!(\partial M_z/\partial H_z)$. Now from the fluctuation dissipation theorem, the magnetic susceptibility is given by:

\begin{eqnarray}
\chi_z &=& \frac{1}{kT} \triangle^2(M_z) = \frac{1}{kT}(\langle
M_z^2\rangle-\langle M_z\rangle^2) \label{put} = \frac{1}{kT} \left(
\sum_{i,j=1}^N \langle s_z^i s^j_z\rangle-\left\langle \sum_{i=1}^N
s_z^i \right\rangle^2 \right),
\end{eqnarray}
where $\triangle^2(M_z)$ is the variance of magnetization. This definition of susceptibility gives a measure of the correlation function, $\langle s_z^i s^j_z\rangle$, which could have a magnetic field dependence and one can measure the field dependence of susceptibility and find the field dependence of correlation function. \\

However, in order to measure the generalized magnetic susceptibility, as a function of magnetic field (and not zero field susceptibility), the same procedure could be followed, but the value obtained is the derivative of magnetization with respect to field. This approximation mentioned in the equation above, breaks down above the quantum critical point, where, $B~>~J/k_B$ and the assumption of isotropy of space breaks down. The validity of this assumption holds for levels of B which is smaller or comparable with J such that the system is still to a large extent isotropic. In addition to the isotropy of space, there are other subtleties involved in the extraction of relevant quantum information from the susceptibility data and is indicated below.

We have obtained a theoretical relationship for the field dependence of susceptibility and hence the correlation function and is given in Eq. \ref{susc_fld}.

\begin{equation}
\chi(H)= (g^2 \mu^2_B N \beta)(0.75 + \frac {\frac{1}{4}e^{-(\frac{J}{4}+H)\beta} + \frac{1}{2}e^{-(\frac{J}{4})\beta}- \frac{3}{2} e^{\frac{3J}{4}\beta} + \frac{1}{4}e^{-(\frac{J}{4}-H)\beta} }{e^{-(\frac{J}{4}+H)\beta} + 2e^{-(\frac{J}{4})\beta}+ 2 e^{\frac{3J}{4}\beta} + e^{-(\frac{J}{4}-H)\beta}}), \label{susc_fld}
\end{equation}

where, $\beta=\frac{1}{k_{B}T}$.

Fig. \ref{susc_fit} shows the fit to the experimental susceptibility data as a function of field, collected at 2~K. Here we have shown the plots only up to 4 Tesla. The experimental value of susceptibility as a function of field (open circles) matches the theoretical value (bold curve). However, above the quantum critical field, there is a down turn in the data (see fig. \ref{susc_fit}) as using the experimental technique employed here to measure the susceptibility, one is not measuring the magnetic field dependence of the correlation function, $\sum_{i=1}^N \langle s_z^i s^j_z\rangle$. In this limit, these two are not the same, i.e. $\chi(H)=\frac{1}{kT} \left(\sum_{i,j=1}^N \langle s_z^i s^j_z\rangle-\langle \sum_{i=1}^Ns_z^i \rangle^2 \right) \ne \partial M_z/\partial H$. The susceptibility measurement measures only the latter and this was also verified by taking the derivative of the magnetization data ($\partial M/\partial H$).

The only way to ensure that we are really measuring $\chi(H) = \sum_{i,j=1}^N \langle s_z^i s^j_z\rangle$ and not the derivative $\!\partial M/\partial H$, is to measure the correlation function using inelastic neutron scattering measurements. \\

\begin{figure}[t]
\includegraphics[width=0.6\columnwidth,clip]{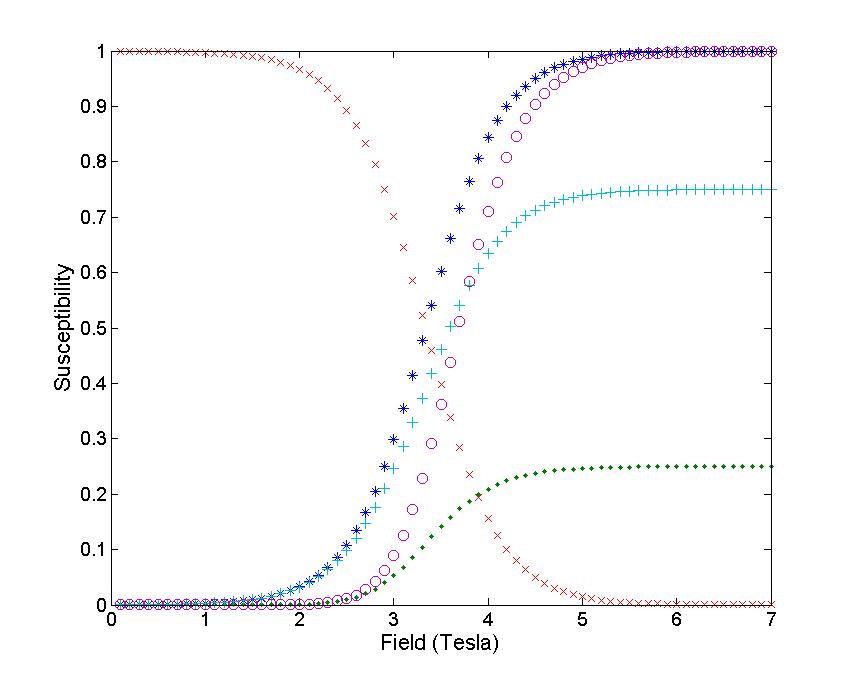}
\caption{Theoretical susceptibility as a function of field at temperature T~=0.5~K, using the correlation function $\left(\sum_{i,j=1}^N \langle s_z^i s^j_z\rangle-\langle \sum_{i=1}^Ns_z^i \rangle^2 \right)$ ('+'). The term $\left\langle \sum_{i=1}^Ns_z^i \rangle^2 \right)$ at T~=0.5~K is shown in the figure (dotted curve). The curve '*' shows the susceptibility as a function of field at T~=0.5~K, obtained using $\left(\sum_{i,j=1}^N \langle s_z^i s^j_z\rangle\right)$. This is also obtained by adding the first two curves. The curves with the symbol 'o' represents 'P' and 'x' represents 'Q' at T~=2~K, where the latter uses the susceptibility plotted in the curve '*'.} \label{Suscept_Z1Z2}
\end{figure}

Another important point is that in a dimer like antiferromagnetic system, since the average spin, $\langle s_z^i\rangle=0$, at least below a critical field, where the ground state is the singlet Bell state $|\Psi^-\rangle$. However, above a critical field where the quantum phase transition takes place, the ground state crosses over to a triplet like state, which is a product state and  $\langle s_z^i\rangle \ne 0$. In the definition of correlation function, $\left(\sum_{i,j=1}^N \langle s_z^i s^j_z\rangle-\left\langle \sum_{i=1}^Ns_z^i \right\rangle^2 \right)$, the second term will start contributing to the susceptibility, after the quantum critical point.
Thus, when we use the term $\left(\sum_{i,j=1}^N \langle s_z^i s^j_z\rangle-\langle \sum_{i=1}^Ns_z^i \rangle^2 \right)$ in the susceptibility to plot the P+Q as a function of field and temperature, it shows an asymmetry either side of the quantum critical point even at the lowest temperature. This is shown in fig. \ref{PQ_3D}(a), which gives the theoretical 3D plot of P+Q as a function of field and temperature. This is what one would obtain if we consider the experimental susceptibility in constructing $Q \!\equiv\! 1-\frac{kT\bar{\chi}}{Ns}$. This asymmetry is in disagreement with what $Wiesniak~et.~al.$ \cite{Vlatko2} obtained. If we ignore the second term in constructing susceptibility, i.e. if we consider only $\left(\sum_{i,j=1}^N \langle s_z^i s^j_z\rangle\right)$, then we obtain the 3D ``P+Q" plot as obtained by $Wiesniak~et.~al.$ and is shown in fig. \ref{PQ_3D}(b).

Before constructing the 2D ``P+Q" as a function of temperature, one needs to demonstrate the different 2D curves of susceptibility mentioned above as a function of temperature. Fig. \ref{Suscept_Z1Z2} shows the susceptibility as a function of field at T~=0.5~K, constructed by using the correlation function, $\left(\sum_{i,j=1}^N \langle s_z^i s^j_z\rangle-\langle \sum_{i=1}^Ns_z^i \rangle^2 \right)$ (curve ``+''). The term $\left\langle \sum_{i=1}^Ns_z^i \rangle^2 \right)$ is also plotted (dotted curve) at the same temperature as a function of field. The sum of these two curves is also shown in this figure (curve ``*"). Next we show the non-local quantity ``Q'' as a function of magnetic field at T~=0.5~K (curve ``x"). This was constructed using the modified susceptibility ($i.e.$ using the curve ``*'' mentioned above). The local term ``P" is also plotted in this figure (curve ``o") at the same temperature as a function of magnetic field. We have chosen T~=0.5~K for all the 2D plots for consistency as we wanted to demonstrate the difference in the susceptibilities owing to different terms in the correlation function. This temperature T~=0.5~K was chosen to bring out the differences clearly as both "P" and "Q" are quite sharp at such a reasonably low temperature.

\begin{figure}[t]
\includegraphics[width=1.0\columnwidth,clip]{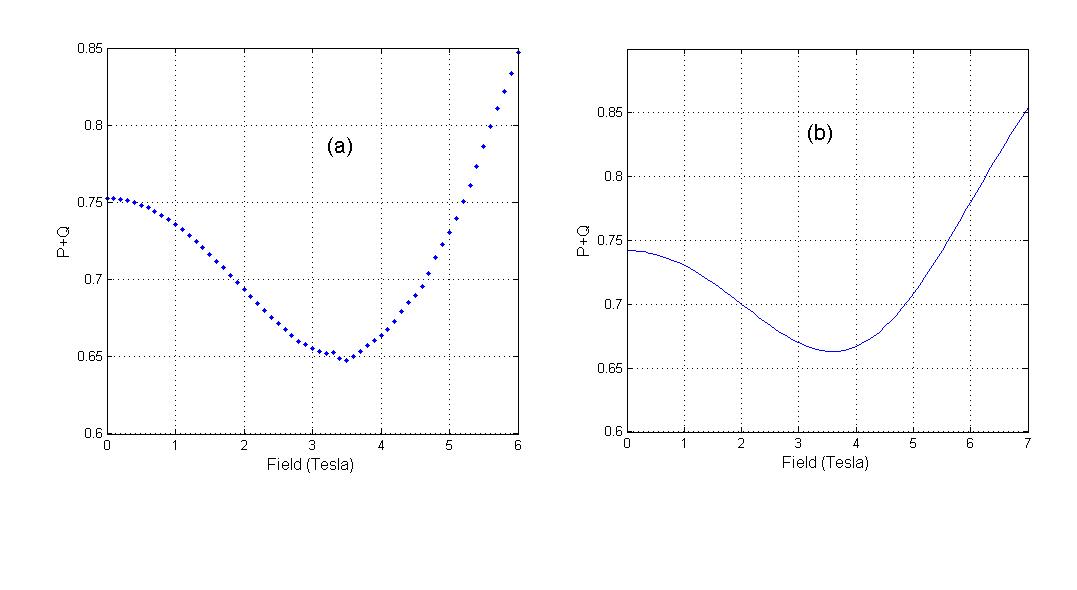}
\caption{(a) Plot of experimental ``P+Q" as a function of magnetic field at T~=2~K. Here the ``Q" used was obtained considering the corrected susceptibility. (b) Plot of theoretical ``P+Q" as a function of magnetic field at T~=2~K. Here the ``Q" used was obtained considering the corrected susceptibility.
} \label{P+Q_2K_Theo}
\end{figure}

Next, we show the theoretical plot of ``P+Q" as a function of magnetic field at T~=2~K in fig. \ref{P+Q_2K_Theo}(b). It shows a dip at the quantum critical point which is between 3.5 Tesla and 4 Tesla. To match the experimental value of ``P+Q" with the theoretical value at T~=2~K, we consider the fit to the experimental susceptibility shown in fig. \ref{susc_fit}. Here in constructing ``Q", we consider the susceptibility data till the magnetic field where the theoretical curve matches the experimental data and then extrapolate the data up to 7 Tesla, using the fitted curve. Since the experimental data measures the term $\left\langle \sum_{i=1}^Ns_z^i \rangle^2 \right)$ as well, we need to add this term to match the experimental data with the theoretical value of $Wiesniak~et.~al.$. This plot is shown in fig. \ref{P+Q_2K_Theo}(a). The local property "P" was considered using the experimental magnetization data at T~=2~K using the formulae, $P\equiv\frac{\langle\vec{M}\rangle^2}{N^2s^2}$. One can clearly see the dip in the value of ``P+Q" in fig. \ref{P+Q_2K_Theo}(a), corresponding to the QCP at around 3.5 Tesla. This suggests that the ``P+Q" prescription proposed by $Wiesniak~et.~al.$ indeed is a detector of quantum phase transition. Our experimental results also demonstrate the quantum information sharing and complementarity between the two observables, susceptibility and magnetization. The ``P+Q" inequality given in Eq. \ref{comp} is not violated in our system and the minimum value is indeed reached at QPT.

\section{Conclusion}
We have studied a Heisenberg spin chain compound known to exhibit a dimer or alternative dimer like characteristics. The system has an antiferromagnetic coupling and therefore exhibits an entangled ground state. This entanglement which was extracted from the susceptibility data decreases when the temperature is varied from the lowest attainable value (1.8 K) and vanishes around 5~K, where the antiferromagnetic correlations die down. We also studied the magnetic field dependence of the magnetization and susceptibility and they both fit reasonably well to an alternating dimer model. We extracted the magnetic field dependence of entanglement and observed that the concurrence decreases with increase in field, owing to an increase in occupation of the excited states in comparison to the antiferromagnetic ground state. At the critical field where, there is a crossover between the ground and excited states, bringing about a change in symmetry of the ground states, a quantum phase transition occurs. We explored the complementarity in the two observables, susceptibility and magnetization, which capture the non-locality and local nature of correlations in this spin system. Following an established prescription, the susceptibility and magnetization were suitably combined to form operators ``P" and ``Q", such that the minimum of the sum ``P+Q" when plotted as a function of magnetic field captures the quantum phase transition. From this measure (``P+Q"), we could capture the QPT as a function of field at 2K. This shows that ``P+Q" is a good measure for detecting quantum phase transitions, even at non-zero temperatures. This could have tremendous implications in quantum information processing as entanglement is known to scale close to QPT from theoretical results. This scale invariance of entanglement coupled with the ``P+Q" measure could be harnessed to execute quantum communications in spins chains. In future, we are making an effort to explore the quantum phase transition at lower temperatures. We also intend to do specific heat measurements to capture other aspects of entanglement and QPT in this system. \\

The authors would like to thank the Ministry of Human Resource Development,
Government of India, for funding. CM would also like to thank Prof P.K. Panigrahi and Philipp Gegenwart for discussions.

\end{document}